\documentclass[apj]{emulateapj}

\usepackage{subfigure}

\shorttitle{CONSTRAINTS ON VHE $\gamma$-RAYS FROM PSR B1951+32}

\begin{document}


\title{Constraints on the steady and pulsed Very high energy gamma-ray
emission from Observation of PSR B1951+32 / CTB 80 with the MAGIC
Telescope}


\author{
 J.~Albert\altaffilmark{a},
 E.~Aliu\altaffilmark{b},
 H.~Anderhub\altaffilmark{c},
 P.~Antoranz\altaffilmark{d},
 A.~Armada\altaffilmark{b},
 C.~Baixeras\altaffilmark{e},
 J.~A.~Barrio\altaffilmark{d},
 H.~Bartko\altaffilmark{f},
 D.~Bastieri\altaffilmark{g},
 J.~K.~Becker\altaffilmark{h}
% ...
}
 \altaffiltext{a} {Universit\"at W\"urzburg, D-97074 W\"urzburg, Germany}
 \altaffiltext{b} {Institut de F\'\i sica d'Altes Energies, Edifici Cn., E-08193 Bellaterra (Barcelona), Spain}
 \altaffiltext{c} {ETH Zurich, CH-8093 Switzerland}
 \altaffiltext{d} {Universidad Complutense, E-28040 Madrid, Spain}
 \altaffiltext{e} {Universitat Aut\`onoma de Barcelona, E-08193 Bellaterra, Spain}
 \altaffiltext{f} {Max-Planck-Institut f\"ur Physik, D-80805 M\"unchen, Germany}
 \altaffiltext{g} {Universit\`a di Padova and INFN, I-35131 Padova, Italy}
 \altaffiltext{h} {Universit\"at Dortmund, D-44227 Dortmund, Germany}
% ...

\submitted{Received 2007 February  2; accepted 2007 July 23}
\journalinfo{The Astrophysical Journal, 669:1143-1149, 2007
November 10}



\begin{abstract}
We report on very high energy $\gamma$-observations with the MAGIC
Telescope of the pulsar PSR B1951+32 and its associated nebula,
CTB 80. Our data constrain the cutoff energy of the pulsar to be
less than $32\,$GeV, assuming the pulsed $\gamma$-ray emission to
be exponentially cut off. The upper limit on the flux of pulsed
$\gamma$-ray emission above $75\,$GeV is
$4.3\cdot10^{-11}\,$photons cm$^{-2}$ sec$^{-1}$, and the upper
limit on the flux of steady emission above $140\,$GeV is
$1.5\cdot10^{-11}\,$photons cm$^{-2}$ sec$^{-1}$. We discuss our
results in the framework of recent model predictions and other
studies.
\end{abstract}


\keywords{acceleration of particles --- gamma rays: observations
---- pulsars: individual (PSR B1951+32) --- radiation mechanisms: nonthermal}



\section{Introduction}

It is currently believed that pulsars are among the few objects in
our Galaxy that are candidate sources of ultrarelativistic charged
cosmic rays. Relativistic particles within the magnetosphere emit
$\gamma$-rays at energies up to several GeV in various processes
such as curvature radiation, synchrotron radiation and inverse
Compton (IC) scattering. Thus, observations in the multi-GeV
$\gamma$-ray domain allow one to  study the acceleration sites in
the magnetosphere of a pulsar. Predicted sites where particle
acceleration can take place are, for example, above the polar cap
of the neutron star
\citep[e.g.][]{1978ApJ...225..226H,1982ApJ...252..337D} and in the
so-called outer gap of the magnetosphere
\citep[e.g.][]{1986ApJ...300..500C,1986ApJ...300..522C,1992ApJ...400..629C}.
Furthermore, particle acceleration can take place outside the
magnetosphere in the region where the pulsar wind interacts with
the interstellar medium. If electrons are accelerated in these
shocks, they could give rise to IC-scattered photons from, for
example, the cosmic microwave background, synchrotron radiation,
or thermal origin
\citep{1992ApJ...396..161D,1996MNRAS.278..525A,2003A&A...405..689B}.
\pagebreak

PSR B1951+32 was detected first at radio frequencies by
\cite{1988Natur.331...50K}, and is one of the six rotation-powered
high energy pulsars whose GeV emission was detected by EGRET
\citep{1995ApJ...447L.109R}. Among $\gamma$-ray pulsars, PSR
B1951+32 is the only source observed to emit up to $20\,$GeV with
no  cutoff being evident in the differential energy spectrum. The
spectrum shows a hard spectral index of 1.8 between $100\,$MeV and
$20\,$GeV. The pulsar has an apparent high efficiency
($\sim0.4\%$) of converting its rate of rotational energy loss,
$3.7\times10^{36}\,$ergs$\,$s$^{-1}$, into $\gamma$-rays above
$100\,$MeV (assuming a distance of $2.5\,$kpc to the pulsar).
Moreover, the $\gamma$-ray luminosity at $\sim10\,$GeV is
comparable to that of the Crab pulsar \citep{1995ApJ...447L.109R}.

AS inferred from its rotational parameters, the spin-down age of
PSR B1951+32 is $\sim 10^5\,$yr \citep{atnf},\footnote{See
http://www.atnf.csiro.au/research/pulsar/psrcat.} that is,~about
100 times older than the Crab pulsar. The magnetic field strength
of $4.9\cdot10^{11}\,$G \citep{atnf} is lower than that in most
rotation-powered pulsars. Because of the lower magnetic field,
curvature $\gamma$-rays emitted near the stellar surface, as
predicted in polar-cap models, are less affected by magnetic pair
production. Compared with younger, more strongly magnetized
pulsars, the spectral cutoff energy is thereby shifted to higher
energies, up to a few tens of GeV
(\citet{2001AIPC..558..115H,Baring}; see also \cite{2000bulik} for
a discussion of low-field millisecond pulsars).

On the contrary, if the $\gamma$-rays are emitted in the outer
magnetosphere, as predicted in outer gap models, the potential
drop in the outer gap of PSR B1951+32 is expected to be comparable
to that of young pulsars (see eq.~[12] of \cite{Zhang1997} and
eq.~[2.1] of \cite{1986ApJ...300..500C}). Therefore, the cutoff
energy, which reflects the maximum Lorentz factor of the electrons
or positrons accelerated in the outer gap, is expected to be
around 10\,GeV \citep{hirotani2006b}. Thus, features in the
predicted spectral shape of weakly magnetized pulsars at energies
above 10\,GeV are strongly dependent on the emission altitudes. In
order to discriminate between emission models, PSR B1951+32 is a
prime candidate for observation by ground-based $\gamma$-ray
detectors with low energy thresholds such as the imaging air
Cherenkov telescope MAGIC.

This pulsar is located in the core of the radio nebula CTB 80,
which is thought to be physically associated with the pulsar. In
X-rays the nebula  shows a cometary shape
\citep{2004ApJ...610L..33M,2005ApJ...628..931L}, being confined by
a bow shock that is produced by the pulsar's high proper motion
($240\pm40\,$km s$^{-1}$) \citep{2002ApJ...567L.141M}.
\cite{2005JPhG...31.1465B} predict an over-$200\,$GeV flux from
the nebula at a level of $\sim 4.4\%$ of the Crab's flux, by
assuming that high-energy leptons can accumulate inside the
well-localized nebula for  long periods of time, as observed in
the case of the Crab nebula.

The current tightest constraint on the emission above $100\,$GeV
from the pulsar and its nebula, obtained by the Whipple
collaboration \citep{1997ApJ...489..170S}, puts an upper limit of
$75\,$GeV on the cutoff energy of the pulsed emission and an upper
limit of $1.95\times10^{-11}\,$cm$^{-2}$s$^{-1}$, on the steady
emission above $260\,$GeV. The latter is within a factor $\sim2$
of the prediction of \cite{2005JPhG...31.1465B}.

\begin{table}[htb]
\tiny \def\arraystretch{1.5}
    \centering
        \caption{\label{datasummary1951}Summary of the observations of PSR B1951+32}
       \begin{tabular}{cccccc}\hline\hline
        Date & Rate & ON Time & Extinction & Extinction Scatter &  \\
         (2006)        & (Hz) & minutes & (mag) & (mag) & Selected? \\
        \hline

        Jul 4........ & 164 & 130 &  0.099 & 0.017 &yes\\
        Jul 5........ & 164 & 136 &  0.100 & 0.011 &yes\\
        Jul 6........ & 167 & 105  & 0.088 & 0.014 &yes\\
        Jul 7........ & 176 &  62 &  0.091 & 0.011 &yes\\
        Aug 3 ....... & 151 & 95  &  0.161 & 0.009 &yes\\
        Aug 4 ....... & $\ldots$ & $\ldots$ &  0.266 & 0.045 &no\\
        Aug 23 ..... & 175 & 168  & 0.079 & 0.017 &yes\\
        Aug 24 ..... & 158 & 105 &  0.088 & 0.014 &yes\\
        Aug 25 ..... & 165 & 138 & 0.142 & 0.029 &yes\\
        Aug 26 ..... & 135 & 148 &  0.168 & 0.044 &no\\
        Aug 27 ..... & 167 & 124& 0.140 & 0.042 &yes\\
        Aug 28 ..... & $\ldots$ & $\ldots$ & 0.249 & 0.056 &no\\
        Sep 13 ...... & 147  & 83   &  $\ldots$     &  $\ldots$   & yes  \\
        Sep 14 ...... & 139  &  155   &  0.105     &  0.016 & yes \\
        Sep 15 ...... & 156  &   102  &  0.091  & 0.017   & yes  \\
        Sep 16 ...... & 147  &   125  &   0.095 &  0.013  & yes  \\
        Sep 17 ...... &  149    &   89    & 0.094 & 0.060     & yes  \\
        \hline

       \end{tabular}
       \tablecomments{The extinction coefficients are taken from publicly
available data from the Carlsberg Meridian Telescope, which is
located on the same site as MAGIC.
        The extinction coefficient is for an effective wavelength of
        625nm.}
    \end{table}

In this paper, we present upper limits on the cutoff energy of the
pulsed emission from the pulsar, as well as on the steady and
pulsed very high energy (VHE) fluxes from the region associated
with the radio nebula, resulting from MAGIC telescope observations
that were performed in 2006 July through September. The paper is
structured as follows: After a short introduction to MAGIC and our
data taking and analysis (\S \ref{analysis}), we report on our
search for steady and pulsed emission from PSR B1951+32 (\S
\ref{results}). We close with a discussion of the implications of
our results (\S \ref{discussion}).

\section{Observations and Analysis}
\label{analysis}

The MAGIC (Major Atmospheric Gamma Imaging Cherenkov) Telescope,
see \cite{2004NewAR..48..339L}, is located on the Canary Island of
La Palma (2200 m above sea level, $28.45^\circ$N, $17.54^\circ$W).
MAGIC is currently the largest imaging atmospheric Cherenkov
telescope (IACT), having a 17 m diameter tessellated reflector
dish comprising 964 $0.5\times0.5\,\mbox{m}^2$ diamond-milled
aluminium mirrors. The faint Cherenkov light flashes produced by
air showers are recorded by the telescope camera, which consists
of 577 photomultiplier tubes. Together with the current
configuration of the MAGIC camera with a trigger region of
2.0$^\circ$ diameter \citep{2005ICRC....5..359C}, this results in
a trigger collection area for $\gamma$-rays of about $10^5\,$m$^2$
at small zenith angles.  The effective collection area depends on
the analysis and is $\sim10^4\,$m$^2$ around $60\,$GeV and
increases to $\gtrsim6\cdot10^4\,$m$^2$ beyond $200\,$GeV. At
present, the minimum trigger energy is 50-60\,GeV (at small zenith
angles). The MAGIC Telescope is focused to 10\,km distance
--- the most likely position for a 50\,GeV air shower maximum. The accuracy
in reconstructing the direction of incoming $\gamma$-rays on an
event by event basis (point spread function), is about
0.1$^\circ$, depending on energy and the chosen analysis method. A
source with a $\gamma$-ray flux of $\sim2$\% that of the Crab
Nebula and the same spectral slope can be detected by MAGIC above
$200\,$GeV at a significance level of 5\,$\sigma$ within 50 hours.

 \begin{figure}[htb]
        \centering
\includegraphics*[bb = 0 0 550 365, angle=0,width=\columnwidth]{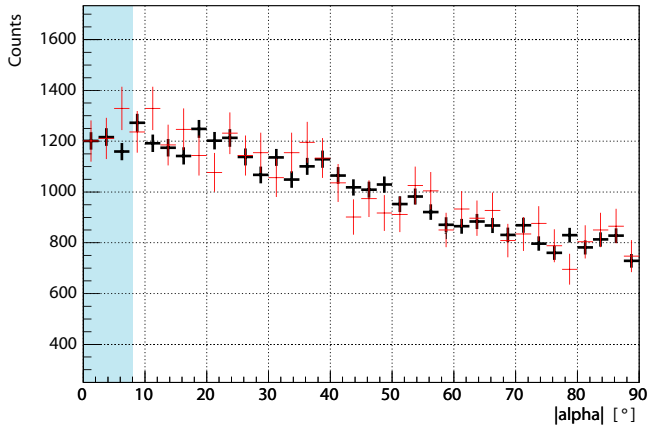}
                        \caption{Distribution of the parameter $|alpha|$ for events $\gtrsim280\,$GeV.
                        The distribution of OFF-source events (\textit{red}) was normalized to the ON-source
                        events (\textit{black})
                        between $20^{\circ}$ and $85^{\circ}$. An excess due to $\gamma$-rays from PSR B1951+32 is expected for
                        $|\mbox{alpha}|<7.5^{\circ}$ (\textit{shaded region}).}
                        \label{alpha_1951}
  \end{figure}

PSR B1951+32 was observed with MAGIC for a total of 17 nights
between 2006 July 4 and September 17. The observations were
performed in the so-called ON/OFF mode; that is, PSR B1951+32 was
observed by directly pointing to it (ON). Three nights were
rejected because of unstable trigger rates due to bad weather. The
background was estimated by observing  at the same range of zenith
angle for 5.8\,hr a suitable region in the sky where no
$\gamma$-ray source is expected (OFF).  In total, 30.7\,hr of data
were processed. The zenith-angle range of the observation was
restricted to between $5^{\circ}$ and $25^{\circ}$, guaranteeing
the lowest possible energy threshold. A summary of the
observations is given in Table \ref{datasummary1951}. This table
also includes the atmospheric extinction coefficients for all
nights, provided by the Carlsberg Meridian Telescope, which is
located at the same site as MAGIC.

Following calibration of the data \citep{2005ICRC....5..375G} and
a tail-cut image cleaning of the events, a Hillas parametrization
algorithm was applied \citep{1985ICRC....3..445H}. The tail cuts
used in the image cleaning were 6 photoelectrons for core pixels
and 4 photoelectrons for boundary pixels. For the generation of
sky maps, we used tail cuts of 10 and 5 photoelectrons. Additional
suppression of pixels containing noise was achieved by requiring a
narrow time coincidence between adjacent pixels ($\sim7\,$nsec).
The hadronic background was suppressed with a multivariate method,
the Random Forest \citep{2001breiman,2004NIMPA.516..511B}, which
uses the Hillas parameters of an event to decide on its so-called
hadronness. The power to suppress hadronic background is energy
dependent and reduced for $\gamma$-ray energies below $150\,$GeV.
As a consequence, the optimal cut in hadronness, which gives the
highest rejection of background while retaining most of
$\gamma$-ray candidates, has to be independently determined for
each energy region. For the analysis of the data presented here,
we used an energy dependent hadronness cut, whose empirical
parametrization was derived from Monte Carlo (MC) studies. An
exception is the sky maps, for which a static hadronness cut was
applied in the event selection. This is justified, as the maps
were produced for energies above $200\,$GeV, where the dependence
of the optimal hadronness cut on energy is small. The method of
random forests is also used to estimate the energy of an event.
Typically, energy resolutions of $\sim25\%$ are achieved on an
event-by-event basis.

\section{Results}
\label{results}

\subsection{Search for Steady Emission}

\begin{figure}[htb]
        \centering
\includegraphics*[bb = 110 29 573 697, angle=-90, width=\columnwidth]{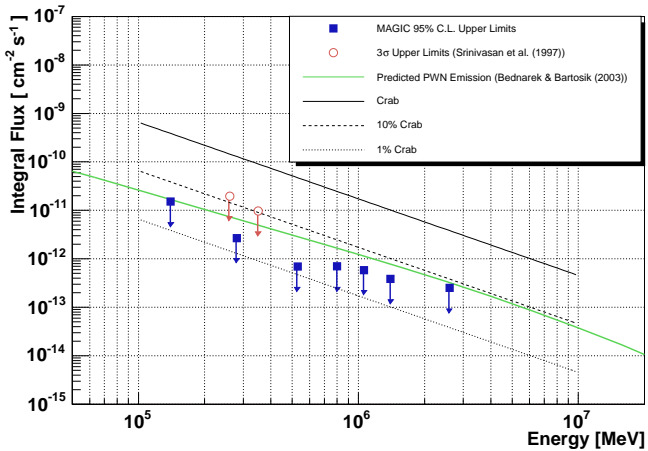}
                        \caption{Integral upper limits (95 \% confidence level) on the steady $\gamma$-ray emission from the direction of PSR B1951+32.
                        For comparison, the $\gamma$-ray flux of the Crab Nebula
                        \citep{2005ICRC....4..163W} is also indicated. }
                        \label{dc_limits}
  \end{figure}

We searched for steady $\gamma$-ray emission of a point source
from the direction of PSR B1951+32 with different analysis
thresholds between $140\,$GeV and $2.6\,$TeV. We define the
analysis threshold as the peak of the energy distribution
    of MC events after cuts. Images of $\gamma$-rays from PSR B1951+32
point with their major axis to the camera center and thus appear
as an excess at small values in the parameter "alpha". Alpha is
the angle between the major axis of the shower image and the
direction determined by the image's center of gravity and the
camera center. In Figure \ref{alpha_1951}, we show the
distribution of $|$alpha$|$ for events with energies
$\gtrsim280\,$GeV. An excess due to $\gamma$-ray emission from PSR
B1951+32 should be visible in the figure for
$|\mbox{alpha}|<7.5^{\circ}$. The results of this analysis and
others with different analysis thresholds are summarized in Table
\ref{summarydc1951}. As no significant signal ($>5\,\sigma$) from
$\gamma$-rays was found, we calculated upper limits on the number
of excess events with a confidence level of 95\% by using the
method of \cite{2005NIMPA.551..493R}. In the calculation of the
limits, a systematic uncertainty on the flux of 30\% was taken
into account. The upper limits on excess events were converted
into integral flux limits by assuming a spectral index of 2.6,
which is similar to the spectral index of the predictions and
other known pulsar wind nebulae (PWNs) such as the Crab Nebula. If
a harder spectrum with index 2.0 is assumed, the flux limits
increase by about 15\% and they decrease by about 40\% if a softer
spectrum with index 4.0 is assumed. The integral flux limits of
$\gamma$-rays are shown in Figure \ref{dc_limits} together with
the measurement of \cite{1997ApJ...489..170S} and the predictions
of \cite{2003A&A...405..689B}.

      \subsection{Search for $\gamma$-Ray Emission in the Vicinity
      of PSR B1951+32}
    We explored the region in the sky around the position
     of the pulsar for a possible extended or displaced emission
    region of $\gamma$-rays. The latter is a likely scenario
    because of the high proper motion of the pulsar. For this
    study, we employed the DISP method of
    \cite{1994APh.....2..137F} with a modified parametrization
    \citep{2005ICRC....5..363D}, which permits the reconstruction of
    the arrival direction of a
     $\gtrsim100\,$GeV $\gamma$-ray with an accuracy of
    $\sim0.1^{\circ}$. Sky maps were produced in different bins of
    energy. In none of the maps was $\gamma$-ray emission found within the
    reconstructed field of view, of $\sim0.6^{\circ}$ radius.

\begin{table*}[htb]
\def\arraystretch{1.1}
\scriptsize
    \centering
        \caption{\label{summarydc1951} Results of the Analysis Searching for
         Steady $\gamma$-Ray Emission from PSR B1951+32.}
       \begin{tabular}{ccccccc}\hline\hline
        \rule[0mm]{0mm}{ 4mm}Analysis Threshold & & & & Significance & Upper Limit, & Flux Upper Limit\\
           \rule[-2mm]{0mm}{ 0mm}      (GeV) & ON Events & OFF Events & Excess Events& ($\sigma$) & Excess Events (95\% C.L.) & (cm$^{-2}$s$^{-1}$)  \\
        \hline

        \rule[0mm]{0mm}{ 4mm}$>140$ .....................................& 37869 & $37933\pm381$& -64 & -0.2 & 792 & $1.5\times10^{-11}$ \\
        $>280$ ..................................... & 3576 & $3740\pm150$\phs & -164 & -1.0 & 196 & $2.7\times10^{-12}$ \\
        $>530$ .....................................& 712 & $777\pm42$\phs & -65 & -1.3 & 54 & $7.0\times10^{-13}$  \\
        $>800$ .....................................& 232 & $231.5\pm22$\phs&0.5 & 0.0 &  55 & $7.0\times10^{-13}$  \\
        $>1060$ ...................................& 101 & $90.6\pm14$&10.4 & \phd0.6 & 45 &  $5.8\times10^{-13}$  \\
        $>1400$ ...................................& 58 & $49.5\pm10.8$&8.5 & \phd0.6 & 35 & $3.9\times10^{-13}$  \\
        $>2600$ ...................................& 17 & $26\pm10$&-9 & \phd-0.9 & 14 &  $2.5\times10^{-13}$
        \\\hline
      \end{tabular}
      \end{table*}

\begin{figure*}[htb]
        \centering

 \subfigure
{\includegraphics*[angle=0,width=8.5cm]{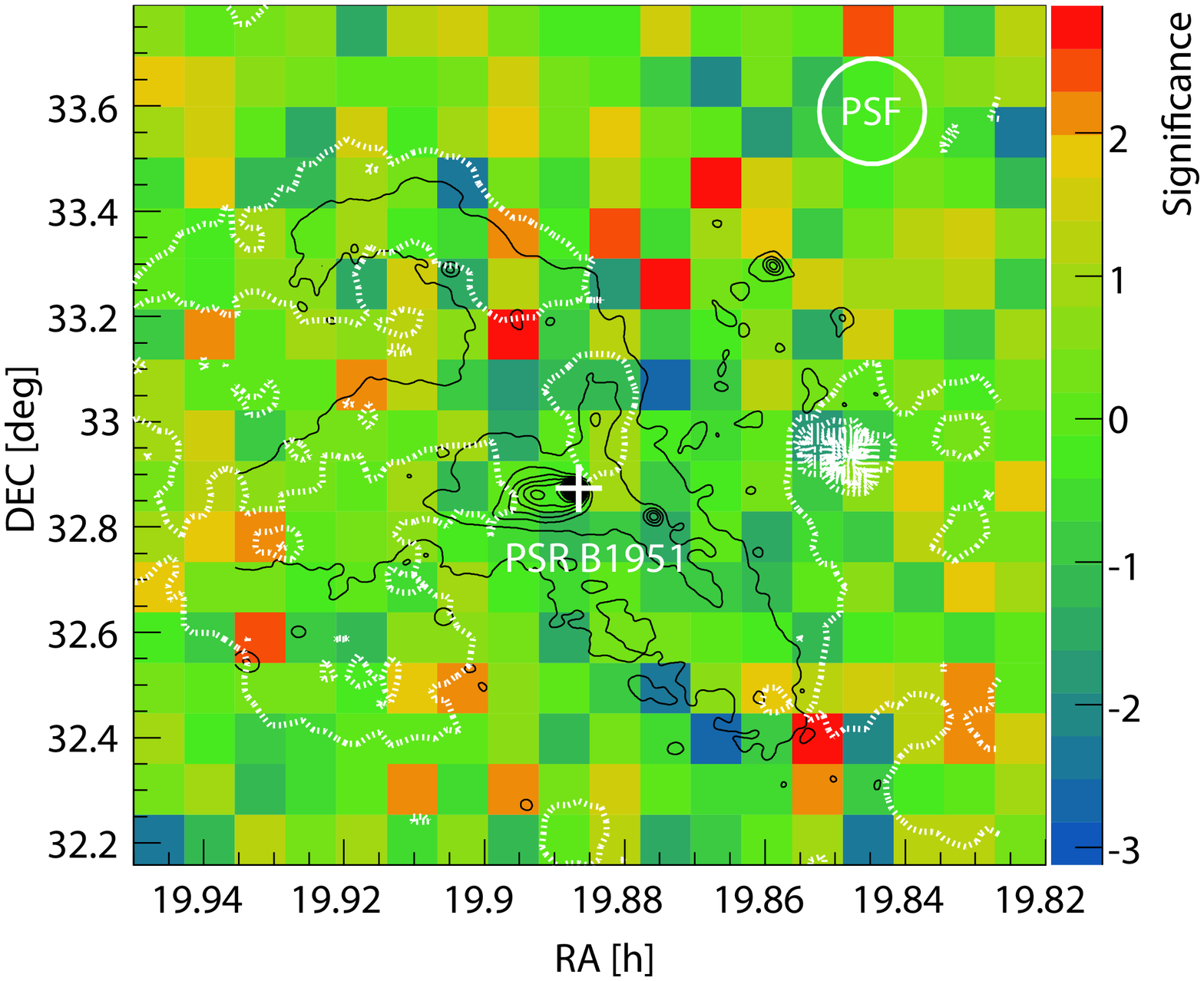}} \hfill
\subfigure{\includegraphics*[angle=0,width=9cm]{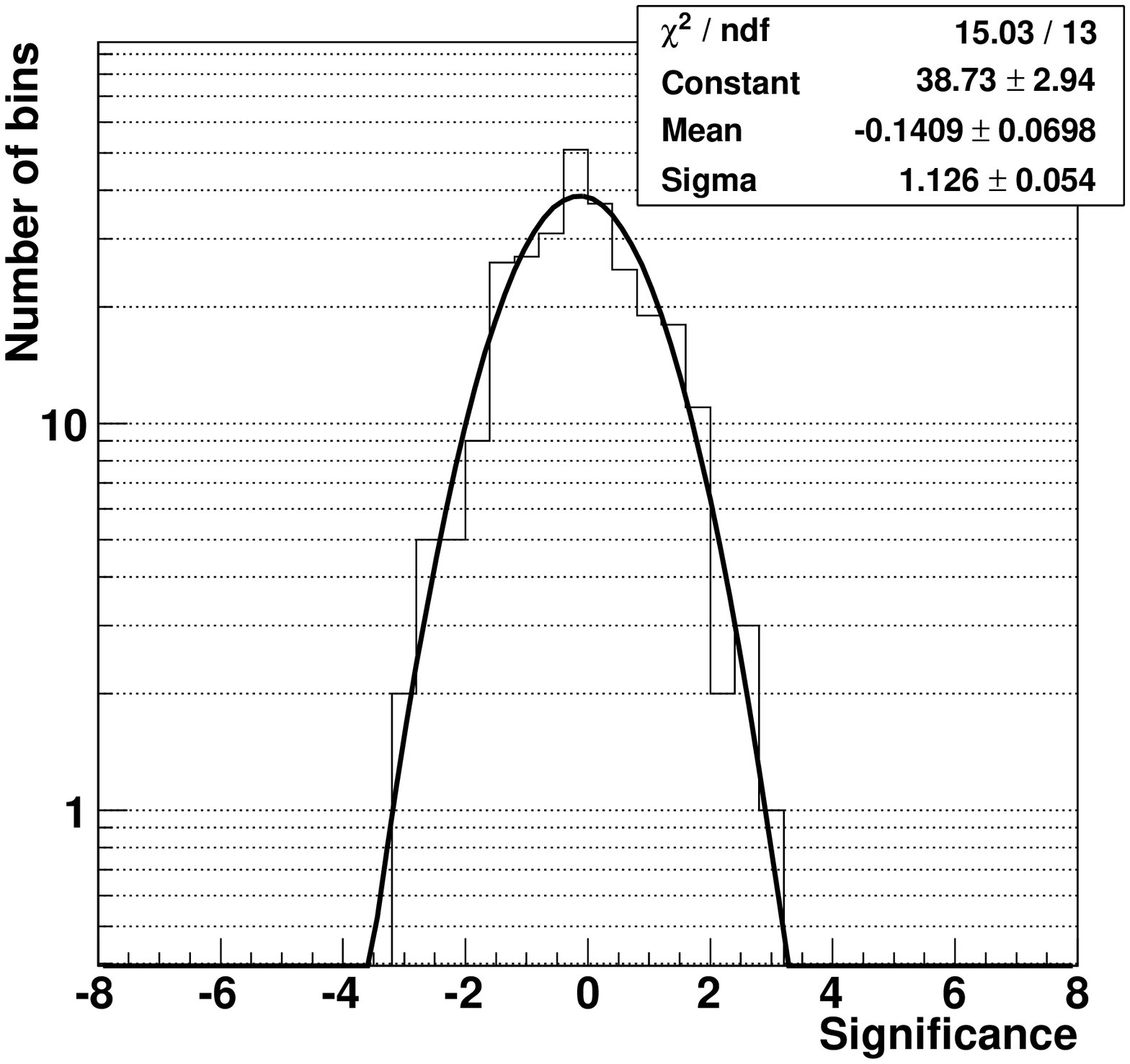}}
                        \caption{Significance of VHE $\gamma$-ray
                        emission from the region around PSR B1951+32. \textit{Left}:
                        Calculated significance of VHE $\gamma$-ray emission $\gtrsim200\,$GeV in bins of
                        \mbox{$0.1^\circ\times0.1^\circ$}.
                        Overlaid in black are contours from radio observations
                        \citep{2003astro.ph.10655C} and in white contours from IR observations \citep{1988Natur.334..229F}. \textit{Right}:
                        Distribution of significances. The
                        distribution is compatible with that of randomly
                        distributed data. \label{1951_map}}
  \end{figure*}

\begin{figure}[htb]
        \centering
\includegraphics*[ angle=0,width=\columnwidth]{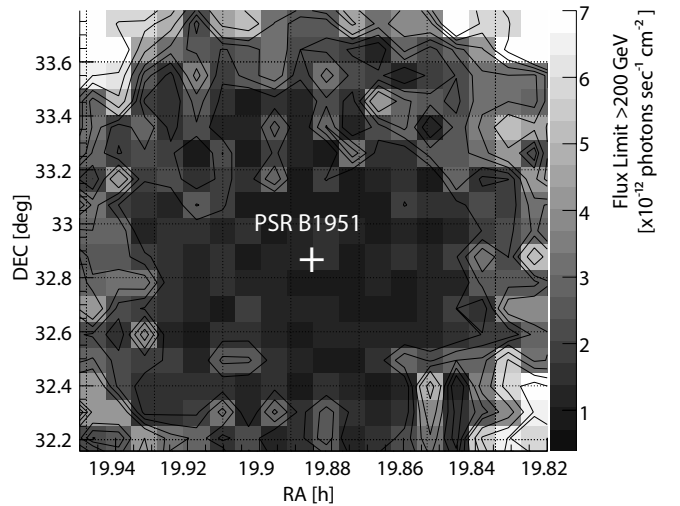}
                        \caption{Upper limits (95 \% confidence level) on the integral $\gamma$-ray emission above $200\,$GeV, calculated in
                        bins of $0.05^\circ\times0.05^\circ$.}
                        \label{dc_limits_map}
  \end{figure}

    The map in Figure \ref{1951_map} (\textsl{left}) shows the significance calculated in bins of
    $0.1^\circ\times0.1^\circ$ for events with energies
    $\gtrsim200\,$GeV. Figure \ref{dc_limits_map} shows a map of the calculated upper
    limits (\mbox{95 \%} confidence level) on the integral flux for the same
    events. The acceptance of the MAGIC camera
    was modelled using the radial dependence of the background rate
    in the camera after event selection. By comparing
    with MC simulations, we confirmed for various angular distances
    from the camera center that the radial dependence of the background rate
    is compatible with the simulated $\gamma$-ray acceptance.

    Following our study, we can exclude steady $\gamma$-ray emission above $200\,$GeV
    at the level predicted by \cite{2003A&A...405..689B}, which we would have
    detected if (1) the emission were originating from within a circle of
    radius $\approx0.4^{\circ}$ centered on the position of the
    pulsar and (2) the apparent emission region was restricted to less than
    $\sim0.3^{\circ}$ in diameter.

\subsection{Search for Pulsed Emission}

    The time of each event (hereafter ``arrival time'') is derived from the time signal of a
    GPS-controlled rubidium clock with a precision of $\sim200$
    ns. Before we searched for pulsed emission from the pulsar, the arrival times
    were
    transformed to the barycenter of the solar
    system with the Tempo timing package by J.~H.~Taylor et al.\footnote{See http://www.atnf.csiro.au/research/pulsar/tempo.} Afterwards, the corrected
    arrival times $t_j$ were folded to the corresponding phase $\phi_j$
    of PSR B1951+32:
    \[
        \phi_j=\nu(t_j-t_0)+\frac 1 2 \dot{\nu}(t_j-t_0)^2+\frac 1 6
        \ddot{\nu}(t_j-t_0)^3
    \]
    where $\nu$, $\dot{\nu}$, $\ddot{\nu}$ and $t_0$ are the values from a contemporary
    ephemeris provided by A.~Lyne (2006, private communication), which is listed in Table
    \ref{ephemeris1951}.
        The analysis chain that was set up to search for pulsed emission was
    previously tested on data from an optical observation of the Crab pulsar
    with the central pixel of the MAGIC camera \citep{2005ICRC....5..367L}.
    Details of the optical observation can be found in F.~Lucarelli et al.~(2007, in preparation).

    We performed a search for pulsed $\gamma$-ray emission from PSR B1951+32
    in five
    differential bins of reconstructed energy between $100\,$GeV and $2\,$TeV.
    To test for periodicity, we applied the Pearson-$\chi^2$ test, the \textsl{H}-Test
    \citep{1989A&A...221..180D}, and a test from
    \cite{1992ApJ...398..146G} (a Bayesian-Test).
    No signature of pulsed emission
    was found in any of the energy intervals. As an example, we give the results
     of the \textsl{H}-Test, which
     yielded
     significances of 0.3, 2.3, 0.6, 0.2 and $1.4\,\sigma$,
    respectively, with increasing energy. The corresponding 95\% confidence level upper
    limits on pulsed emission are shown in Figure
    \ref{1951_pulsar}. The limits were calculated from
    the results of the \textsl{H}-Test \citep{1994ApJ...436..239D} by assuming a duty cycle
    for the pulsed emission of 36\%, which corresponds to the duty cycle
     of PSR B1951+32 at energies above $100\,$MeV
    \citep{1995ApJ...447L.109R}. A spectral slope of 2.6 was assumed in the calculation
    of the upper flux limit. Note that these are upper limits in
    differential bins of energy, whereas the upper limits from
    Whipple \citep{1997ApJ...489..170S} are integral ones,
    which were converted to differential ones assuming
     a spectral shape of 2.6.

    \begin{table}[htb]
    \centering
        \caption{\label{ephemeris1951}Ephemeris of PSR B1951+32} \hspace*{-1cm}
       \begin{tabular}{cc}\hline\hline
       \rule[-2mm]{0mm}{ 6mm}Parameter& Value\\\hline
       \rule[0mm]{0mm}{ 4mm}Position epoch (JD) ......... & 2,450,228.4144 JD\\
       R.A. .................................
       &$19\fh52\fm58\fs27568995$\\
       Decl. ................................ &
       $32\arcdeg52\arcmin40.6824033\arcsec$\\
        Pulsar Epoch (JD) ..........& 2,453,931.724208 JD\\
        $\mathbf \nu$ (Hz) ..............................& 25.29516019929(63)\\
        $\dot{\nu}$ (Hz s$^{-1}$) .......................& $-3.72818(33)\cdot10^{-12}$ \\
        \rule[-2mm]{0mm}{ 0mm}$\ddot{\nu}$ (Hz s$^{-2}$) .......................&$-1.15(25)\cdot10^{-21}$\\\hline

       \end{tabular}
       \tablecomments{From A.~Lyne (2006, private
       communication).\\
         Uncertainties are given in parentheses.}
    \end{table}

    In a second analysis, we searched for pulsed
    emission by selecting events with a
    SIZE$>100\,$photoelectrons\footnote{SIZE is the integrated intensity of a shower image
    after applied tail cuts in units of photoelectrons. It is also a good measure of the
    incident energy for shower impact parameters between $\sim50$ to $120\,$m.},
     that is, events with energies $\gtrsim75$GeV. Again,
    no hint of pulsed emission was found. The \textsl{H}-Test yielded
    1.4, and a $\chi^2$ test 7.2 with 11 degrees of
    freedom. The Bayesian test gave a probability for
    pulsed emission of $2.4\cdot10^{-4}$.

    From the result of the \textsl{H}-Test, we calculated an upper
    limit on the number of excess events (s.~Table
    \ref{1951_pulsarsearch_summarytable}), from which we derived
    an upper limit on the cutoff energy of the pulsed emission in the following
    way:
    The known spectrum of PSR B1951+32 at GeV energies, measured
    by EGRET
    \citep{1996PhDT.........1F}, was multiplied by an exponential
    cutoff and convolved with the effective collection area of the
    telescope. For a given cutoff energy, we then obtained the number of expected
    excess events by multiplying the result with
    the dead-time-corrected observation time.
    The upper limit on the cutoff energy was finally found by
    iteratively changing the cutoff energy until the number of
    expected  excess events matched the upper limit on
    the number of pulsed excess events. With this procedure we obtained
     an upper limit on the cutoff energy
    of $32\,$GeV.
    The measured spectrum of PSR B1951+32 multiplied
    by an exponential cutoff of $32\,$GeV is shown in
    Figure \ref{1951_pulsar} (\textit{red curve}). The analysis threshold,
    $75\,$GeV, is marked by the red arrow in the figure. In case that the rollover of the $\gamma$-ray spectrum is
    superexponential in shape,
    we constrain the cutoff energy to be below 60\,GeV.

\begin{figure}[htb]
        \centering
                    \includegraphics*[bb = 190 30 565 700, angle=-90,width=0.9\columnwidth]{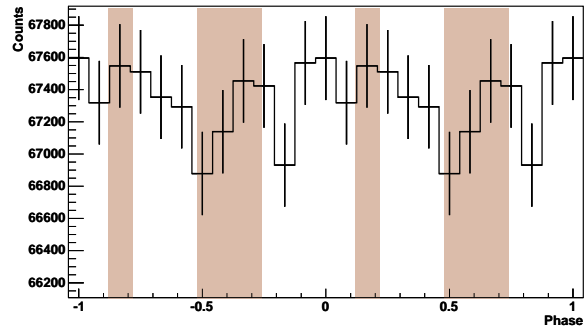}
                        \caption{Pulse phase profile of PSR
                        B1951+32 obtained after selecting events with SIZE $<300\,$photoelectrons. The shaded areas
                        indicate the phase regions in which PSR B1951+32 is emitting at GeV energies \citep{1995ApJ...447L.109R}.}
                        \label{lclower300phe}
   \end{figure}

    As a cross-check, the same analysis was repeated, this time by selecting all events
    with a SIZE $<300\,$photoelectrons, that is, events with energies $\lesssim180\,$GeV.  The resulting pulse phase profile in Figure \ref{lclower300phe}
    shows no evidence for pulsed emission.
    From this analysis, a slightly better upper limit on
    the cutoff energy of $28\,$GeV results.
    The analysis threshold, $60\,$GeV, was lower because
    events with a SIZE below
    100 photoelectrons were also included in the analysis.

\section{Discussion}
\label{discussion}

Theoretical predictions and experimental evidence from lower
energies had been quite favorable for a possible detection of
$\gamma$-ray emission from PSR B1951+32 or its nebula with MAGIC.
Nevertheless, despite the higher sensitivity of this observation
compared to previous ones, no $\gamma$-ray emission was detected.

The upper limits in Figure \ref{dc_limits} on the steady
$\gamma$-ray emission from the PWN surrounding PSR 1951+32 are
below the $\gamma$-ray flux that was predicted by the
time-dependent model of
\cite{2003A&A...405..689B,2005AIPC..745..329B}. Although their
model takes into account the temporal evolution of the nebula (but
not the spatial evolution), the acceleration of leptons and
therefore also the equilibrium spectrum of leptons inside the
nebula still depends on a few free parameters. These parameters,
for example, the density of the medium surrounding the PWN, the
acceleration efficiency of leptons, and the magnetization
parameter of the pulsar wind at the shock region, are not well
constrained by observations.

 \begin{figure}[htb]
        \centering
\includegraphics*[ angle=0,width=\columnwidth]{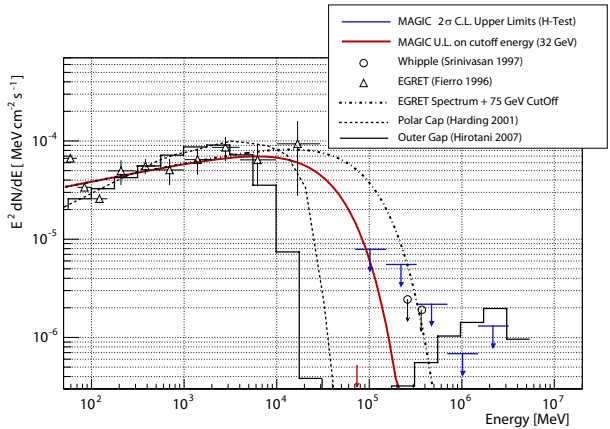}
                        \caption{Results of the analysis in the search for pulsed emission from PSR B1951+32.
                        Upper limits are given with a 95 \% confidence level. The upper limit on the cutoff energy from Whipple is shown as the dot-dashed curve.
                        The upper limit on the cutoff of 32 GeV by MAGIC is shown as the solid red curve. The analysis threshold (75\,GeV)
                        is marked by the arrow on the horizontal axis.}
                        \label{1951_pulsar}
  \end{figure}

Concerning the magnetization parameter, that is, the ratio of the
magnetic energy flux to the particle energy flux,
\cite{2005ApJ...628..931L} have recently estimated the magnetic
field strength of the compact X-ray nebula around PSR B1951+32 to
be $\sim300\,\mu$G, which is larger than the value assumed by
\citeauthor{2003A&A...405..689B}. At the present time it is
therefore clear that the value of the magnetization parameter
$\sigma$ of the pulsar wind has to be much larger than the value
of $\sigma = 10^{-3}$, which \citeauthor{2003A&A...405..689B}
assumed. As a result, the cooling of electrons by synchrotron
radiation is faster and the IC $\gamma$-ray flux is suppressed.
Nevertheless, a hadronic component, as predicted in some models
\citep{2003A&A...405..689B,2006A&A...451L..51H}, which would
dominate if the acceleration efficiency of leptons was low
\citep{2006astro.ph.10307B}, would be below the sensitivity of our
observation.

Another aspect is that the model of
\citeauthor{2005AIPC..745..329B} deals with PWNe that are well
confined by the external medium and pulsars that are, at most,
moving slowly through the interstellar medium (the prototype of
such a nebula is the Crab nebula). Only in such a scenario should
a well-localized $\gamma$-ray source  be expected, whereas when a
pulsar is moving very fast, the $\gamma$-ray emission will be
distributed over a larger volume. In the case of PSR B1951+32,
which is moving with an apparent velocity
\mbox{$v_\mathrm{PSR}=240\pm40$ km s$^{-1}$}
\citep{2002ApJ...567L.141M}, the $\gamma$-ray flux estimated by
\cite{2005JPhG...31.1465B} will be smeared over an area with a
diameter $d$ of at least
\begin{equation}
 d=v_\mathrm{PSR}\tau_\mathrm{PSR}\approx5.3\cdot10^{19}\,\mbox{cm}\approx0.5^\circ,
\end{equation}
assuming an age of the pulsar of $\tau_\mathrm{PSR}=7\times 10^4$
yr and a distance of 2\,kpc. Such an extended emission region
reduces the detection probability with MAGIC. Apart from the
pulsar's motion and the diffusion of leptons, their confinement
and cooling, as well as their injection rate into the interstellar
medium over time, have to be taken into account. These parameters
are unknown, and therefore, their influence on the extension of
the $\gamma$-ray source is difficult to estimate. Assuming Bohm
diffusion in a magnetic field of $3\times10^{-6}\,$G, one
estimates a diffusion length of $\sim13\,$pc for 100\,TeV leptons
during the lifetime of the pulsar \citep{2005JPhG...31.1465B}. In
this case the extension would marginally increase by
$\sim0.1^\circ$ beyond what is expected from  the motion of the
pulsar alone. If the magnetic field distribution is ordered, the
diffusion can be faster and even anisotropic, leading to much
larger emission regions. In this context it is interesting to note
that extended TeV $\gamma$-ray sources associated with displaced
pulsars were recently detected by the H.E.S.S. Collaboration
(e.g., the Vela pulsar \cite{2006A&A...448L..43A} and PSR B1823-13
\cite{2005A&A...442L..25A}).

Considering the $\gamma$-ray emission from the pulsar, we
constrain the cutoff of the pulsed emission to less than $32\,$GeV
if the cutoff is an exponential, which is appropriate when the
$\gamma$-rays are emitted more than a few neutron star radii above
the surface. If photons are emitted at lower altitudes, they are
subject to magnetic pair production, resulting in a stronger
(superexponential) attenuation of the energy spectrum. In the
latter scenario we constrain the allowed range of cutoff energies
to be $\lesssim60\,$GeV.  Considering further that large
uncertainties govern the last spectral point measured by EGRET, it
follows that the allowed energy region where the cutoff resides
can be constrained to lie somewhere between 10 and $30\,$GeV
(exponential cutoff) or up to 60\,GeV (superexponential cutoff).
The narrow allowed range does not leave much freedom for models.
This result and the upper limits from the search in differential
bins of energy are compared in Figure \ref{1951_pulsar} with
theoretical predictions from the polar-cap and the outer-gap
model. In this figure, the dotted line represents the polar-cap
predictions from \citep{2001AIPC..558..115H}, renormalized to the
points of the EGRET spectrum. The thin solid line shows the
spectrum of the latest outer-gap model \citep{hirotani2006b}.

\begin{table*}[htb]
    \centering
    \def\arraystretch{1.4}
        \caption{\label{1951_pulsarsearch_summarytable}Results of the Analysis for Periodicity}
       \begin{tabular}{ccccccc}\hline\hline
       &\multicolumn{4}{c}{\textsl{H}-Test}&&\\
\cline{2-5}
 & & Significance & $2\sigma$ U.L.,&
        $2\sigma$ Flux U.L. &  &\\
         SIZE cut& Result & $(\sigma)$ & Excess Events&(cm$^{-2}$s$^{-1}$) & $\chi^2$ &\textsc{Bayesian}\\\hline
        \textsc{Size} $>100\,$phe .........................& 1.4 & $0.3\sigma$ & 2188 & $4.3\cdot10^{-11}$ & 7.2 & $2.4\cdot10^{-4}$ \\
        \textsc{Size} $<300\,$phe .........................& 3.2 & $1.1\sigma$ &  3388 & $5.0\cdot10^{-11}$ &10.7 &
        $3.6\cdot10^{-4}$ \\\hline
       \end{tabular}
    \end{table*}

In polar-cap models, the cutoff energy is determined by the
attenuation of $\gamma$-rays due to magnetic pair production and
hence by the emission altitude of $\gamma$-rays. As a consequence,
the energy spectrum above the cutoff energy is superexponentially
attenuated. If the emission altitude in the polar cap model shown
in Figure \ref{1951_pulsar} changes from 1 to 2 stellar radii, the
cutoff energy will increase from 20\,GeV to 60\,GeV, which is,
according to our observations, the maximum allowed cutoff energy
for a superexponentially shaped cutoff. On the contrary, in
outer-gap models, the cutoff is determined by the maximum Lorentz
factor of the accelerated positrons and electrons. As a
consequence, the cutoff of the $\gamma$-ray spectrum is smoother,
resulting in an exponential cutoff. If the magnetic field lines
near the light cylinder are straighter than assumed for the
outer-gap spectrum in Figure \ref{1951_pulsar}, the predicted flux
below 60\,GeV will increase.

For more precise predictions of the cutoff energy in polar-cap
models, multidimensional and self-consistent electrodynamics have
to be examined from first principles, whereas a three-dimensional
magnetic field configuration has to be investigated in the
outer-gap model. Assuming that these improvements in theory will
be achieved in the near future, measurements  with higher
statistics around $10\,$GeV, for example, by GLAST, or
measurements by future ground-based experiments with lower
thresholds than MAGIC, for example, MAGIC II or the Cherenkov
Telescope Array, will be needed in order to distinguish between
models.

The predicted IC flux at TeV energies in the outer-gap model
(Figure \ref{1951_pulsar} \textit{solid black line}) appears to be
inconsistent with our upper limits. Nevertheless, it must be noted
that the IC flux is obtained by assuming that all the
magnetospheric soft photons illuminate the equatorial region of
the magnetosphere in which the gap-accelerated positrons are
migrating outwards. Therefore, the predicted IC flux as a function
of energy specifies an upper boundary to the possible pulsed TeV
emission. The open poloidal magnetic field lines could have a
single-signed curvature within 1.8 light-cylinder radii, as the
solution of the time-dependent force-free electrodynamics of an
oblique rotator indicates \citep{2006ApJ...648L..51S}. If this is
the case, soft photons emitted inside the light cylinder along the
convex magnetic field lines will not efficiently illuminate the
magnetic field lines, which are slightly convex even outside the
light cylinder. As a result, the predicted IC flux at TeV energies
will be significantly reduced. This problem will be solved in
future when the self-consistent gap electrodynamics
\citep{hirotani2006a,hirotani2006b} and the three-dimensional
force-free electrodynamics are combined.

We are grateful for the preparation of the ephemeris of PSR
B1951+32 by Andrew Lyne, thus enabling us to perform the pulsed
analysis. Alice Harding was so kind as to provide us with her
polar-cap predictions of PSR B1951+32. We also would like to thank
the Instituto de Astrofisica de Canarias for the excellent working
conditions at the Observatorio del Roque de los Muchachos, in La
Palma. The support of the German Bundesministerium f\"ur Bildung
und Forschung and the Max-Planck-Gesellschaft, the Italian
Istituto Nazionale de Fisica Nucleare, the Spanish Comisi\'on
Interministerial de Ciencias y Tecnolog\'ia, ETH Research Grant TH
34/04 3, and grant 1P03D01028 from the Polish Ministerstwo Nauki i
Informatyzacji are gratefully acknowledged.

\bibliographystyle{plainnat}
\bibliography{}




\end{document}